\documentclass[prl, 10pt,twocolumn, floatfix, groupedaddress, aps, longbibliography, nofootinbib]{revtex4-2}

\usepackage{times, mathrsfs, amsmath, amsfonts, graphics, graphicx, cancel, color, amsthm, bbm, mathtools, amssymb,cases,physics}

\usepackage{amsmath,amsthm,amsfonts,graphicx,xcolor,times,xfrac,booktabs,mathtools,enumitem,xr,subfigure,amssymb,bbm,verbatim,appendix,placeins, physics}

\usepackage[unicode=true,bookmarks=true,bookmarksnumbered=false,bookmarksopen=false,breaklinks=false,pdfborder={0 0 1}, backref=false,colorlinks=true,citecolor=red]{hyperref}
\usepackage{orcidlink}

\usepackage{bm}
\usepackage{dcolumn,algorithm,algpseudocode}

\makeatletter
\theoremstyle{plain}

\theoremstyle{plain}

\theoremstyle{plain}

\theoremstyle{remark}
\newtheorem*{rem*}{\protect\remarkname}
\theoremstyle{plain}

\theoremstyle{plain}

\theoremstyle{definition}

\theoremstyle{plain}
\newtheorem*{thm*}{\protect\theoremname}
\theoremstyle{plain}
\newtheorem*{lem*}{\protect\lemmaname}

\providecommand{\propositionname}{Proposition}
\providecommand{\theoremname}{Theorem}
\providecommand{\lemmaname}{Lemma}
\providecommand{\remarkname}{Remark}
\providecommand{\conjecturename}{Conjecture}
\providecommand{\definitionname}{Definition}
\providecommand{\corollaryname}{Corollary}
\allowdisplaybreaks

\def\ket#1{\vert{#1}\rangle}

\def\BraVert{e.g.,roup\,\mid\,\bgroup}

\def\bbkk#1#2{\langle\!\langle{#1}\vert{#2}\rangle\!\rangle}

\def\Tr#1{\mbox{Tr}\left[\,{#1}\,\right]}
\def\Var#1{\mbox{Var}\left[{#1}\right]}

\DeclareMathOperator{\diag}{diag}

\newcommand{\Id}{\mathbbm{1}}

\def\HiLi{\leavevmode\rlap{\hbox to \hsize{\color{yellow!50}\leaders\hrule height .8\baselineskip depth .5ex\hfill}}}

\begin{document}

\title{Post-processing optimization and optimal bounds for non-adaptive shadow tomography}

\author{Andrea Caprotti\orcidlink{0000-0002-4404-2216}}
\email{andrea.caprotti@univie.ac.at}
\affiliation{University of Vienna, Faculty of Physics, Vienna Center for Quantum Science and Technology (VCQ), Boltzmanngasse 5, 1090 Vienna, Austria}
\affiliation{University of Vienna, Vienna Doctoral School in Physics, Boltzmanngasse 5, 1090 Vienna, Austria}

\author{Joshua Morris\orcidlink{0000-0002-1022-7976}}
\email{joshua.morris@univie.ac.at}
\affiliation{University of Vienna, Faculty of Physics, Vienna Center for Quantum Science and Technology (VCQ), Boltzmanngasse 5, 1090 Vienna, Austria}
\affiliation{University of Vienna, Vienna Doctoral School in Physics, Boltzmanngasse 5, 1090 Vienna, Austria}

\author{Borivoje Daki\'c\orcidlink{0000-0001-9895-4889}}
\email{borivoje.dakic@univie.ac.at}
\affiliation{University of Vienna, Faculty of Physics, Vienna Center for Quantum Science and Technology (VCQ), Boltzmanngasse 5, 1090 Vienna, Austria}
\affiliation{Institute for Quantum Optics and Quantum Information (IQOQI), Austrian Academy of Sciences, Boltzmanngasse 3, 1090 Vienna, Austria.}

\begin{abstract} 
\noindent 
Informationally overcomplete POVMs are known to outperform minimally complete measurements in many tomography and estimation tasks, and they also leave a purely classical freedom in shadow tomography: the same observable admits infinitely many unbiased linear reconstructions from identical measurement data. We formulate the choice of reconstruction coefficients as a convex minimax problem and give an algorithm with guaranteed convergence that returns the tightest state-independent variance bound achievable by post-processing for a fixed POVM and observable. Numerical examples show that the resulting estimators can dramatically reduce sampling complexity relative to standard (canonical) reconstructions, and can even improve the qualitative scaling with system size for structured noncommuting targets.

\end{abstract}

\maketitle

\section{Introduction}
Quantum state tomography~\cite{dariano2003} remains crucial for the characterization of quantum states in multipartite systems, but is hindered in practicality by the prohibitive resource requirements~\cite{odonnell2015,haah2017,aaronson17,kueng2014}.
Shadow tomography~\cite{aaronson17}, and the subsequent adaptation with classical shadows ~\cite{preskill2020,morris2020}, have been proposed as an alternative to characterizing multiple properties of the state simultaneously, following a ``measure first, ask questions later'' paradigm. 
While many adaptations akin to this spirit~~\cite{hong-ye2023, vermersch2020,preskill2021,kraus2021,brierly2021,flammia2021,mezzacapo2022,miyake2021,jiang2023,babbush2023,bryan2024,bu2024,benedetti2023,bertoni2023,koh2022,fava2023,taylor2023,albert2024,nakaji2023,goldstein2022,guehne2023} have been proposed to further improve the computational and experimental overhead, these mostly consider the ``measure first" aspect, aiming to optimize the demand on quantum operations~~\cite{serbyn2022, anshu2021, preskill2023, guehne2022, lu2021, kulik2021, borregaard2024, khemani2023, dakic2022, glos2022, ringbauer2022, vermersch2023}.
Only recently has there been a growing interest concerning the ``ask questions later'', with the realization that there are different strategies to answer the same question from previously sampled data~\cite{caprotti2024,innocenti2023,malmi2024,fischer2024}. 
This freedom in data processing is used to adapt the measurement analysis based on the specific task in order to reduce estimator variance, and constitutes an indirect way of reducing the overall degree of quantum information processing demanded. 

This work expands on the concept of \emph{post-processing optimization} and determining optimal estimators based on the target by introducing a robust convex optimization algorithm for constructing the estimator itself, forming a task-adaptive design problem that can be solved after data acquisition. 
Once a measurement scheme has been chosen and corresponding measurements have been sampled, we show that any \emph{remaining degrees of freedom in the problem live entirely within the post-processing space}. 
One is then free to choose among infinitely many coefficient vectors $\vec{x}$ that satisfy $O=\sum_j x_j E_j$ for some observable $O$ without modifying the underlying experiment (for the case of overcomplete measurements). 
From classical statistics bounds, the variance of this estimator determines the amount of copies of the state to be prepared and sampled: it therefore represents the ideal figure of merit for an optimization protocol to determine the ideal construction of the estimator.

We thus formulate a procedure to identify \emph{optimal variance as a minimax convex problem with guaranteed convergence}, thereby producing the \emph{tightest non-adaptive, state-independent} variance bound attainable by classical post-processing for a fixed measurement scheme. 
Our numerical examples indicate that variance-optimal post-processing can change not only prefactors but also the \emph{scaling} of sampling complexity with system size, including regimes with pronounced qualitative improvements over canonical estimators~\cite{d'ariano_2007}.
In particular, we identify regimes where the variance bound exhibits sub-exponential scaling, in contrast to the exponential behavior predicted by canonical estimators.

The paper is structured as follows: Section \ref{sec:post-processing} is dedicated to the role of post-processing in linear estimation, where in particular in Section \ref{sec:state-of-the-art} we describe the current attempts at post-processing optimization, in Section \ref{sec:classical_shadows} we highlight the reconstruction freedom in the context of the classical shadow protocol and in Section \ref{sec:optimization_problem} we present the algorithm which uses the identified freedom to determine the optimal post-processing procedure for a given linear estimation task. Then, relevant results are introduced: in Section \ref{sec:plane_proj} we show the optimal estimation of product observables with a restricted POVM, which presents an exponential improvement compared to the canonical reconstruction; moreover, in Section \ref{sec:pauli_sum} we present the case of the sum of local observables on a spin chain. Finally, Section \ref{sec:outlook} gives an outlook on the possible extensions of the current algorithm to a wider range of post-processing tasks.

\section{Post-processing optimization}\label{sec:post-processing}
\subsection{State-of-the-Art}\label{sec:state-of-the-art}

In linear tomography, the experimental layer fixes a measurement scheme -typically a POVM $\{E_j\}_{j=1}^n$- while the final estimate of a target quantity is produced by \emph{classical} data processing. 
The central observation behind \emph{post-processing optimization} is that, once the POVM is fixed and outcomes have been sampled, the mapping from the observed frequencies $\hat{f}_j$ into an estimator $\hat{o}=\sum_{j=1}^n \hat{f}_j\,x_j$ generally allows a nontrivial freedom, as the coefficients $x_j$ are not unique as long as the set of POVM effects are linearly dependent (overcomplete). 
In other words, overcompleteness induces a ``gauge freedom'' in post-processing which can be exploited without any change to the experiment. 

The relevance of optimizing this classical layer predates classical shadows. 
In the seminal works on linear estimation by D'Ariano~\cite{d'ariano_2001, d'ariano_2007}, the problem is framed explicitly as the choice of \emph{data-processing functions} (reconstruction coefficients) for POVMs and it is shown how one can reduce estimation noise by optimizing the processing of redundant measurements within the family of unbiased reconstructions.
This line of work established that \emph{``how we reconstruct''} is part of the experimental design space, even when the hardware is fixed.
A key message for full-state estimation was later sharpened by Zhu~\cite{zhu2014}: informationally overcomplete POVMs can yield a quantitative advantage over minimal informationally complete measurements. 
The further relevant point is that the target-agnostic ``canonical" estimator does not appear to be optimal, with optimality being heavily dependent on the figure of merit (e.g. estimator variance). 
The measurement-frame formulation developed in~\cite{guehne2022,innocenti2023} collects and further hones these ideas, also bringing the classical shadow protocol~\cite{preskill2020, morris2020} under such unified framework.
The standard classical-shadows estimator corresponds in fact to the canonical choice of reconstruction, which is natural and broadly applicable but generally non-optimal for any given target. 

Only recently has overcompleteness systematically begun to be exploited as an effective optimization resource~\cite{caprotti2024,malmi2024,fischer2024}. These works emphasize the complementary aspects:
(i) explicit parameterizations of the homogeneous freedom in the inversion map and observable-dependent optimization leading to sharp improvements and scaling separations \cite{caprotti2024},
(ii) data-driven (empirical) optimization of duals tailored to target tasks, including practical strategies to avoid bias when the same data is used for tuning and estimation~\cite{malmi2024},
and (iii) systematic investigations of families of optimized approximate dual frames -including marginal constructions and regularized empirical duals- aimed at achieving improvements under realistic computational constraints \cite{fischer2024}.
Taken together, these results establish that ``measure first, ask later'' should be complemented by ``{\bf \emph{process} later, but process \emph{optimally}}.'' 

In the classical-shadows context, much of the literature uses the shadow norm~\cite{preskill2020} as a convenient, state-independent upper bound controlling sample complexity.
While powerful, such bounds may be loose for a given observable/POVM pair and may not reflect the best achievable performance within the unbiased linear family. 
This motivates treating the \emph{variance itself} as the primary figure of merit and optimizing post-processing with respect to it. 
Besides being operationally direct, a variance-based objective interfaces naturally with robust concentration techniques (e.g. via \ median-of-means) that yield high-confidence guarantees under weaker tail assumptions than those implicitly built into canonical shadow-norm analyses.

\subsection{Post-processing as an estimation resource}\label{sec:classical_shadows}
The core of shadow tomography is the construction of a compressed classical description of an unknown quantum state. Rather than the exact reconstruction of the density matrix, the intent is to reproduce the behavior of the state for a set of characterizing properties. 
By sampling multiple copies of a quantum state with a fixed measurement scheme, the resultant distribution of obtained outcomes may be used to estimate expectation values of target observables. 
The information of the state is therefore contained in the set of measurement outcomes and can be extracted in post-processing based on the relevant properties of choice. 

In the generalization of the classical shadows protocol~\cite{flammia2021,guehne2022} the measurement scheme for a $d$-dimensional system is described by a given POVM. 
Its effects are positive semi-definite observables $\{E_j\}_{j=1}^n$, $d\times d$ matrices that span a subspace $\mathcal{V}_D$ of dimension $D$ of the Hilbert-Schmidt space.
The POVM then forms a frame to express any operator $A\in \mathcal{V}_D$ together with a corresponding \textit{dual frame}, composed by a set of operators $\{\eta_j\}_{j=1}^n$, such that
\begin{equation}\label{eq:dual_frame}
    A = \sum_{j=1}^n \bbkk{E_j}{A} \, \eta_j 
    = \sum_{j=1}^n \bbkk{\eta_j}{A} \, E_j
\end{equation}
where the inner product in the Hilbert-Schmidt space $\bbkk{A}{B} = \Tr{A^\dagger \,B}$.
The classical shadow tomography protocol uses these dual operators as estimators for the reconstruction of any density matrix in the spanned subspace.
The classical shadows are constructed by randomly sampling a POVM and recording, out of $K$ total measurements, the number of occurrences $k^{(j)}$ of each outcome $E_j$; the relative frequencies $\hat{f}_j = k^{(j)}/K$ are then used as weights for the resultant estimator:
\begin{equation}\label{eq:classical_shadow}
    \hat{\rho} 
    \,=\, 
    \frac{1}{K}\sum_{j=1}^n k^{(j)}\,\eta_j 
    \; = \; 
    \sum_{j=1}^n \,\hat{f}_j\, \eta_j.
\end{equation}
This object does not necessarily represent a positive-semidefinite density matrix, but it is designed to reproduce exactly the underlying state in expectation: $\mathbb{E}[\hat{\rho}] = \rho$.
Consequently it may be used as a replacement of $\rho$ for the estimation of its properties, commonly the expectation value of an observable $O$:
\begin{equation}\label{eq:shadow_exp_val}
    \hat{o} = \Tr{O\,\hat{\rho}} = \sum_{j=1}^n \hat{f}_j \Tr{O\,\eta_j} 
\end{equation}
As the number of measurement increases, the relative frequencies approach the actual probability distribution $p_j = \Tr{E_j \rho} = \bbkk{E_j}{\rho}$: by using the dual property of the dual effects $\eta_j$ from~\eqref{eq:dual_frame}, the classical estimator $\hat{o}$ thus converges to the actual expectation value:
\begin{equation}\label{eq:shadow_est_convergence}
    \hat{o} 
    \overset{\hat{f}_j\to p_j}{\xrightarrow{\hspace{25 pt}}}
    \sum_{j=1}^n \Tr{O\, \bbkk{E_j}{\rho}\,\eta_j} = \Tr{O\,\rho} = \ev{O}.
\end{equation}
The estimator $\hat{o}$ converges to the true expectation value following the central limit theorem; the sample size $K$ required to approximate $\ev{O}$ to an accuracy $\varepsilon$ can be obtained using Chebyshev's inequality, which yields 
\begin{equation}\label{eq:scaling_sample_size}
    K = \mathcal{O} \left( \frac{\Var{\hat{o}}}{\varepsilon^2}\,\cdot \log \delta^{-1} \right),
\end{equation}
where $\delta$ is the tolerance level considered. The convergence and concentration can be further elevated to exponential bounds by the median-of-means technique~\cite{preskill2020}, essentially making variance the key figure of merit.
Given the estimation is all done in post-processing, $M$ different observables can be simultaneously estimated from the same measurement sample, with just an additional factor of $\log M$ for the sample size $K$ to guarantee the same level of accuracy $\varepsilon$. 
These bounds are analyzed in depth in Appendix \ref{app:sample_size}.

There exists a ``canonical'' choice of estimators~\cite{d'ariano_2007}, which represents the optimal solution for a fully unbiased target-agnostic estimation. 
This set of estimators is used as benchmark in the rest of this work; an in-depth description of how it may be obtained is given in Appendix \ref{app:canonical_inv}.
However, the choice of estimators for the construction of the shadow is usually not unique: given an overcomplete POVM, where $n>D$, there exist infinite equivalent choices of estimators $\{\eta_j\}_{j=1}^n$ which act as dual frame, as introduced in \eqref{eq:dual_frame}.
Given the set of measurement outcomes, the extraction of the relevant information in post-processing can be performed in multiple equivalent ways: by adapting the choice of estimators to the particular target observable, one can therefore reduce the overall resources required for the state characterization~\cite{caprotti2024}. 

\subsection{Minimax optimization}\label{sec:optimization_problem}

As introduced in the previous section, the required sample size $K$ depends on the variance of the estimator, $\hat{o}$.
Since the probability distribution is not known beforehand, to plan accordingly the actual measurement procedure it is convenient to look at the worst-case scenario, that is to consider the maximal variance attainable, for any given property. 
The optimal set of estimators for a fixed target observable is therefore characterized as the one which guarantees the minimal upper bound on variance for any unknown state.
The optimization problem can therefore be formalized as a \emph{convex \texttt{minimax} problem}, minimizing over the set of estimators whilst maximizing over all available states:
\begin{equation}\label{eq:opt_problem_est}
	\begin{split}
        \Var{\hat{o}} 
    	&= \mathbb{E}[\hat{o}^2] - \mathbb{E}[\hat{o}]^2 \\
    	&\leq 
    		\min_{\eta_j : \text{\scriptsize{d.f. of} } E_j}
            \max_{\rho: \text{ \scriptsize{state}}} 
            \,\sum_{j=1}^n \Tr{\rho \,E_j} \left(\Tr{O\,\eta_j}\right)^2 \\
         & \quad - \left( \sum_{j=1}^n \Tr{\rho \,E_j} \,\Tr{O\,\eta_j}\right)^2 \\
         & \leq \|O\|^2_{\text{shadow}}
    \end{split}
\end{equation}
where the constraint on the operators $\eta_j$ is for them to be a valid dual frame as defined in \eqref{eq:dual_frame}.
The optimal bound on variance is itself upper bounded by the \textit{shadow norm} $\|O\|^2_{\text{shadow}}$, the figure of merit introduced in \cite{preskill2020} and commonly used in shadow tomography to estimate the measurement sample size.

Rather than handling $n$ different $d\times d$ matrices $\eta_j$, it is more convenient to consider the coefficients $x_j = \Tr{O\,\eta_j}$.
Since $\{ \eta_j \}_{j=1}^n$ represent a valid dual frame for the POVM, these coefficients correspond to the decomposition of the target observable $O$ in terms of the effects $\{ E_j \}_{j=1}^n$, as shown in \eqref{eq:dual_frame}: $\sum_j x_j E_j = O$.
The optimization problem can thus be formalized as 
\begin{equation}\label{eq:opt_problem_coef}
    \begin{split}
    	\min_{x_j \in \mathbb{C}}
    	\max_{\rho\in\mathcal{R}} \quad&
    		f(\vec{x},\rho) = \Var{\hat{o}\,\big|\,\vec{x},\rho}
            \\
         \text{s.t.} \quad& \sum_{j=1}^n x_j E_j = O \\
                     & \rho \succ 0       \\
                     & \Tr{\rho} = 1
    \end{split}
\end{equation}
where the bilinear target function $f(\vec{x},\rho)$ is the variance of the target observable $A$ decomposed by the coefficients $\vec{x}$ for fixed state $\rho$ and $\mathcal{R}$ denotes the set of density matrices (positive semi-definite and with unitary trace).

Since the subspaces considered are all finite-dimensional, it can be safely assumed that all coefficients $x_j$ are finite, therefore the norm of each coefficient is bounded by some constant $|x_j|\leq B$. 
The minimization can thus be restricted to the optimization of the tensor circle $\mathbb{B} = \{\vec{x}\in\mathbb{C}^n \,\big| \; |x_j|\leq B \;\forall j = 1\ldots n \} \subset\mathbb{C}^n$.
As shown in Appendix \ref{app:conc_conv}, the function $f(\vec{x},\cdot) : \mathbb{B}\to\mathbb{R}$ for fixed state is convex in the variable $\vec{x}$; instead, for fixed set of coefficients the function $f(\cdot,\rho) : \mathcal{R}\to\mathbb{R}$ is concave.
Given the set of valid density matrices $\mathcal{R}$ is compact and convex, the minimax theorem \cite{simons_1995} applies for this optimization problem.
This firstly implies that a global solution to \eqref{eq:opt_problem_coef} exists, as a saddle point of $f(\vec{x}, \rho) = \Var{\hat{o}\,\big|\,\vec{x},\rho}$.
Secondly, it also implies that the minimization and maximization can be inverted:
\begin{equation}\label{eq:min_max}
    \min_{x_j \in \mathbb{B}} \max_{\rho\in\mathcal{R}} f(\vec{x}, \rho) =
     \max_{\rho\in\mathcal{R}} \min_{x_j \in \mathbb{B}}f(\vec{x}, \rho)
\end{equation}
The problem remains equivalent, but it is helpful as an analytical dependence of the coefficients for fixed state is easily achievable.
We therefore present the minimization and maximization problem separately.

\subsubsection{Minimization over post-processing coefficients}\label{sec:var_min}
First of all, we notice that, given the constraint $\sum_j x_j E_j = O$, the expectation value for a fixed density matrix is the same for any choice of $\vec{x}$.
This implies that the choice of coefficients $\vec{x}^{\,\star}$ that minimizes variance is completely equivalent to the solution for the first term of the variance:
\begin{equation}\label{eq:min_opt}
    \begin{split}
        \min_{x_j \in \mathbb{B}}  \quad  f(\vec{x},\bar{\rho}) & = 
        \min_{x_j \in \mathbb{B}} \vec{x}^{\,\dagger} \, P(\bar{\rho})\,\vec {x} \\
        \text{s.t.} \quad & \sum_j x_j E_j = O
    \end{split}
\end{equation}
where we introduce the diagonal matrix $\bar{P} = P(\bar{\rho}) = \diag{(p_j)}=\diag{(\Tr{\bar{\rho} \,E_j})}$ of elements of the probability distribution for fixed state $\bar{\rho}$. \footnote{It is relevant here to underline how the probability distributions considered cannot be generalized to the set of valid $n-$element probability distributions $\vec{p}$ s.t. $p_j\leq 0,\,\sum_j p_j = 1$: not every possible probability distribution corresponds to an existing density matrix.
Therefore, from now on, when a probability distribution $\vec{p}$ is considered, it will always be intended as $\vec{p}=\vec{p}(\rho)$.}
This implies that the set coefficients which solve \eqref{eq:min_opt} minimizes both the variance and the shadow norm.

For a condensed expression we introduce, as in~\cite{caprotti2024}, the POVM coefficient matrix $R$, whose rows contains the coefficients to reconstruct the POVM effects in terms of the basis elements: $E_j = \sum_k R_{jk} B_j$.
The constraint on coefficients $\vec{x}$ can be simply expressed as $R^\top \vec{x} = \vec{o}$, where the entries $o_k = \ev{O,B_k} = \Tr{O\,B_k}$ represent the coefficients of the target observable expressed in terms of the same basis.

By assuming that $p_j \neq 0\,\forall j$, the minimization is guaranteed an analytical solution:
\begin{equation}\label{eq:opt_coef_state}
    \begin{split}
        \vec{x}^{\,\star} =
        \vec{x}(\bar{\rho})  
        & = 
        \bar{P}^{-1} \, R\,\left(R^\top\,\bar{P}^{-1}\, R\right)^{-1} \vec{o} \\
        & = L^\star(\bar{\rho}) \,\vec{o},
    \end{split}
\end{equation}
where the state dependence is given by the diagonal matrix $\bar{P} = P(\bar{\rho})$.
Explicit calculations and observations can be found in Appendix \ref{app:opt_coef}.

In \eqref{eq:opt_coef_state} we have introduced the $n\times D$ coefficient matrix $L^\star = L^\star(\bar{\rho}) $ to distinguish the dependence of the coefficients from the state and from the observables.
This shows how, for any observable, the choice of optimal coefficients $\vec{x}^{\,\star}$ actually depends on the underlying state, not on the observable itself.
This result is equivalent to the well-known construction of the optimal dual frame for single-shot variance~\cite{d'ariano_2007,zhu2017,innocenti2023}, where the classical estimator $\hat{o}$ is obtained from the full set of dual frame elements $\{\eta_j^{\,\star}\}_{j=1}^n$ rather than through the simplified coefficient decomposition from~\eqref{eq:opt_coef_state}.

\subsubsection{Maximization over states}\label{sec:var_max}
The analytical solution to the minimization problem determines an explicit dependence of the optimal set of coefficients on the state.
The optimization function thus reduces to the concave function $F(\rho) = f\left(\vec{x}^{\,\star}(\rho),\rho\right) : \mathcal{R}\to \mathbb{R}$ in terms of density matrices.
The problem of determining the solution for the tightest upper bound on variance therefore reduces, for any observable, to the maximization of a concave function on the convex set of valid density matrices 
\begin{equation}\label{eq:max_problem_states}
    \begin{split}
        \max_{\rho\in\mathcal{R}} \quad &
    		f(\vec{x}^{\,\star}(\rho), \rho) = \Var{\hat{o}\,\big|\,\vec{x}^{\,\star}(\rho),\rho}
            \\
         \text{s.t.} \quad & \rho \succ 0       \\
                     	   & \Tr{\rho} = 1.
    \end{split}
\end{equation}
 Even if lacking an analytical solution, this shows that shadow tomography generally admits a scaling improvement through a classical post-processing procedure which relies on the solution of a \emph{convex} optimization problem.
 The convexity of \eqref{eq:max_problem_states} implies the problem admits an efficient numerical treatment: the convergence to a solution is guaranteed even by relying on simple gradient descent methods, as the ones used to find the results presented in this work. The schematic representation of the post-processing optimizations is shown in Fig.\ref{fig:sum_pauli_results} a).

\section{Results}\label{sec:results}
The solution to~\eqref{eq:max_problem_states} represents the tightest bound possible to the variance of the estimator $\hat{o}$ obtained from random sampling of the POVM describing the measurement scheme.
Even if the sampled distribution $\hat{f}$ does not exactly match the expected probability distribution $\vec{p}\,(\rho)$, any reconstructed state from this partial distribution is still guaranteed to have a smaller variance that the worst-case-scenario.

We now compare the optimal results with other reconstruction methods to gauge the actual benefit achievable with this post-processing procedure.
In this section a selection of relevant examples of tight upper bound of target observables.
Together with the estimate of the required sample size scaling, we also compare the optimal value of variance with the maximal variance obtained using the canonical estimators, to underline the improvement which can be achieved.
The numerical data produced in this study are available online \cite{caprotti_zenodo2026}.

\subsection{Product observables}\label{sec:plane_proj}

\begin{figure*}
\includegraphics[width=0.98\textwidth]{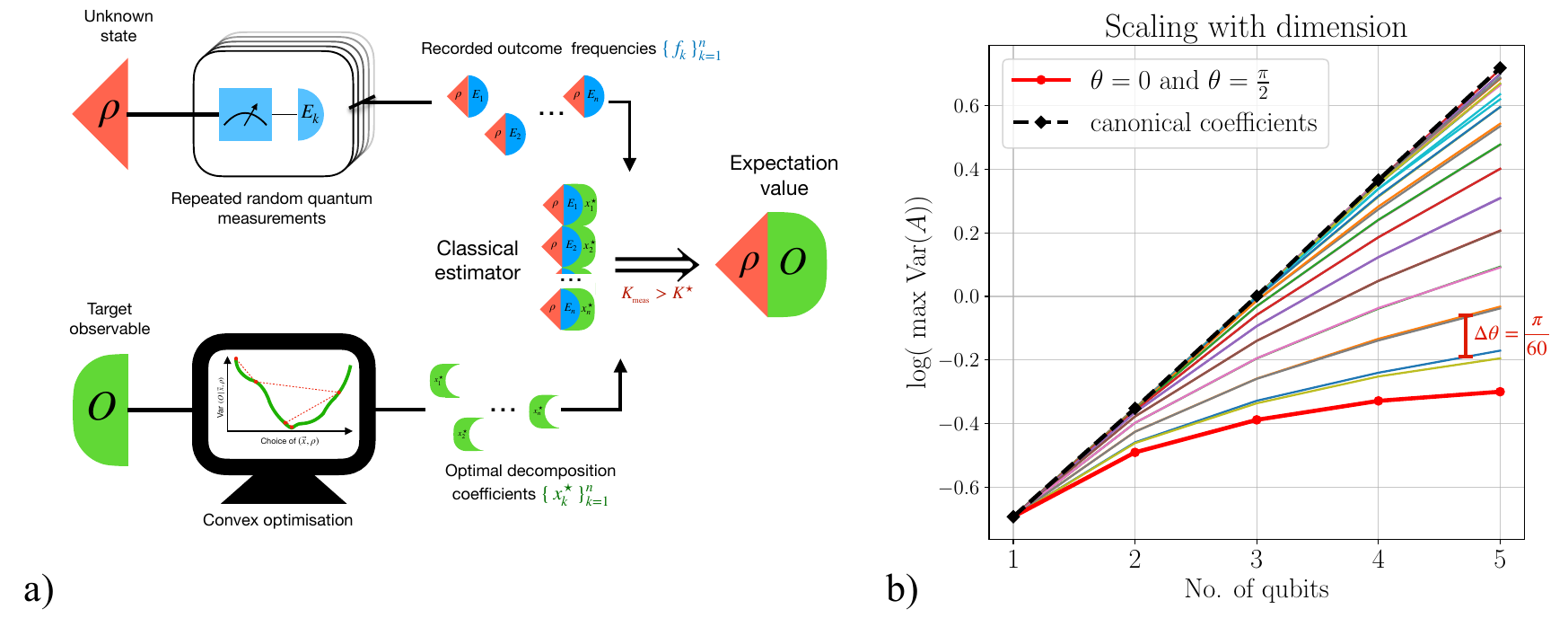}
\caption{ a) \textbf{Post-processing optimization.} 
 The panel schematizes the shadow tomography procedure with optimized post-processing: the unknown state is sampled $K$ with a fixed POVM and the relative frequencies of outcomes $\{\,f_k\}_{k=1}^n$ are recorded; these frequencies are combined with decomposition coefficients $\{x_k^{\,\star}\}_{k=1}^n$ of the target observable in terms of the same POVM, which can be obtained through a classical convex optimization scheme.
The resulting classical estimator thus approximates the actual expectation value $\ev{O}$.
b) \textbf{Optimized variance of product observables.} The optimal upper bound on variance for tensors of single-qubit projectors constrained on the prime meridian of the Bloch sphere in terms of a single angle parameter $\theta\in[0,\pi]$ (see main text for details). 
The panel compares the scaling of the optimal variance (in log scale) with number of qubits for $\theta = \frac{s}{30}\cdot \frac{\pi}{2}$, where the index $s$ assumes all integer values $1\leq s\leq 30$: each run presents and increment of $\Delta \theta = \frac{\pi}{60}$.
The scaling slope varies continuously between the sub-linear optimal case, for $\theta = 0 \wedge \frac{\pi}{2}$ (red line), and the exponential worst case scenario, for $\theta = \frac{\pi}{4}$, which coincides with the scaling obtained from the canonical coefficients (black dotted line).
}
\label{fig:plane_proj}
\end{figure*}

The first example considered is the set of projectors on a restricted subspace. 
The measurement scheme considered is a composed by a subset of the usual Pauli projectors POVM, in particular constituted by the eigenstates of the $X$ and $Z$ Pauli operators normalized to satisfy the requirement $\sum_k^4 E_k = \Id$:
\begin{equation}\label{eq:flat_povm_def}
    E_k = \frac{1}{4}\left(\Id +\,(-1)^k \,\hat{\sigma}_{\lfloor \frac{k}{2}\rfloor} \right), 
\end{equation}
where $\hat{\sigma}_1 = X$ and $\hat{\sigma}_2 = Z$.
For a single qubit state, the subspace spanned by this POVM $\mathcal{V}_{XZ}$ corresponds to the prime meridian of the Bloch sphere.
Since the dimension of this subsystem is $d=3$ and the POVM is constituted by $n=4$ effects, the optimization scheme depends on a single degree of freedom, making it arguably the simplest realistic scenario to be considered.
For multiple qubits, the POVM is taken as $N$-folded tensor product of the single-qubit effects:
\begin{equation}\label{eq:tensor_effects}
    E_{i_1 \dots i_N} = \bigotimes_{k=1}^N E_{i_k}. 
\end{equation}
We set single-qubit projectors (pure states of $\mathcal{V}_{XZ}$), as parametrized by the angle $\theta\in[0,\pi]$ \footnote{In our analysis, the observable angle parameter $\theta$ restricted  between the pole, $\theta = 0$, and the equator, $\theta = \frac{\pi}{2}$, due to the symmetry of the subspace}:
\begin{equation}\label{eq:plane_projectors}
    \begin{split}
    \ket{\psi\,(\theta)} 
    &= 
    \cos{\frac{\theta}{2}}\ket{0} + \sin \frac{\theta}{2}\ket{1} \\
    O(\theta) &= \dyad{\psi(\theta)}{\psi(\theta)},
    \end{split}
\end{equation}
and define an $N$-qubit target observable as $O^{(N)}(\theta)=O(\theta)^{\otimes N}$. 

For the sake of comparison, we shall firstly briefly comment of the canonical estimator for $O^{(N)}(\theta)$ which is being studied together with the optimality analysis in detail in Ref. \cite{caprotti2024}. Namely, when inverting equation $O^{(N)}(\theta)=\sum_{i_1\dots i_N}x^{(N)}_{i_1\dots i_N}E_{i_1\dots i_N}$, the canonical estimator takes the product form
\begin{equation}\label{eq:local coeff}
    \begin{split}
       x^{(N)}_{i_1\dots i_N}&=x_{i_1}\dots x_{i_N},~\mathrm{or}\\
       \vec{x}^{\,(N)}&=\vec{x}^{\,\otimes N},
   \end{split}
\end{equation}
where the coefficients $\vec{x}=(x_1,x_2\dots)$ can be found by the standard inversion procedure ~\cite{d'ariano_2007}. 
We refer to the decomposition \eqref{eq:local coeff} as \textit{local decomposition}, which arises for canonical estimators and product target observables~\cite{caprotti2024}. 
Instead of optimizing local coefficients $x_i$ (as done for the shadow norm figure of merit in~\cite{ caprotti2024}), we provide a full optimization of global coefficients $x^{(N)}_{i_1\dots i_N}$ by the algorithm outlined in the previous section. 
The results of the numerical optimization are shown in Fig.\ref{fig:plane_proj} b). 

As expected, for the single qubit case, the optimization of the variance does not provide any additional improvement compared to the standard case -the solution provided by the canonical coefficient matrix inverse is already optimal.
However, the distinction becomes more and more apparent as the dimension increases.
For $\theta=\frac{\pi}{4}$ the canonical choice appears to be already optimized, presenting the exponential scaling expected for local measurements~\cite{preskill2020}.
For every other value of the $\theta$ parameter, sooner or later the line of the upper bound starts to deviate from the canonical scaling, thus highlighting how the optimized solution does lead to a reduced upper bound on variance.
What is most interesting to notice is that the optimization leads to an improvement also on the \textit{speed} of the scaling: in particular, in the best case scenario of $\theta = 0$, that is projector on the eigenstate of $Z$\footnote{for symmetry of the results, the same behavior is also achieved for projectors on eigenstates of $X$, for $\theta = \frac{\pi}{2}$} the optimal scaling appears to be sub-linear, in what is effectively an exponential improvement compared to the standard procedure.
For every other case, the scaling slope appears to change continuously based on the parameter $\theta$ between the worst case (black triangles) and the best case (red circles) scenario.

The sub-exponential scaling obtained in all but one particular case is very promising, as it shows the possibility of recovering an exponential advantage in the sample size by only relying on post-processing in the choice of how to construct the classical estimator in the shadow tomography scheme.
In general, in the interval $\theta \lesssim \frac{\pi}{10} \vee \theta \gtrsim \frac{2}{5}\pi$ the scaling appears sub-linear: given the continuous dependence of the slope on the parameter $\theta$, it can be expected that there exists a measurement scheme which would guarantee this kind of scaling improvement.

The possibility of observing sub-exponential scaling can be explained by the fact that the optimal variance of tensor solutions is not found as the product of local solutions.
Unlike the case of the optimal shadow norm -as also discussed in the previous work Ref.~\cite{caprotti2024}- the optimal solution of maximal variance \emph{does not necessarily follow the product structure} in \eqref{eq:local coeff}, even though both measurements and target observables are of the product (tensor) form.
As our numerical simulation shows, this makes post-processing optimization richer for significant improvement of sampling complexity, but also harder to fully optimize, as the post-processing space (number of free parameters) grows exponentially with the number of systems.

While this particular example is not very informative for practical applications, as there exist much more efficient and direct fidelity estimation methods, in its simplicity it shows how a consistent and relevant improvement in the scaling can actually be found.
Even though there is still little insight on the characterization of this optimal state, or even of the particular subset of $\mathcal{V}_D$ where to expect the solution to come from, it still suggests that entangled coefficients (global) might present optimal solutions with larger improvement compared to the canonical reconstruction.

\subsection{Sum of local observables}\label{sec:pauli_sum}

\begin{figure*}
\includegraphics[width=\textwidth]{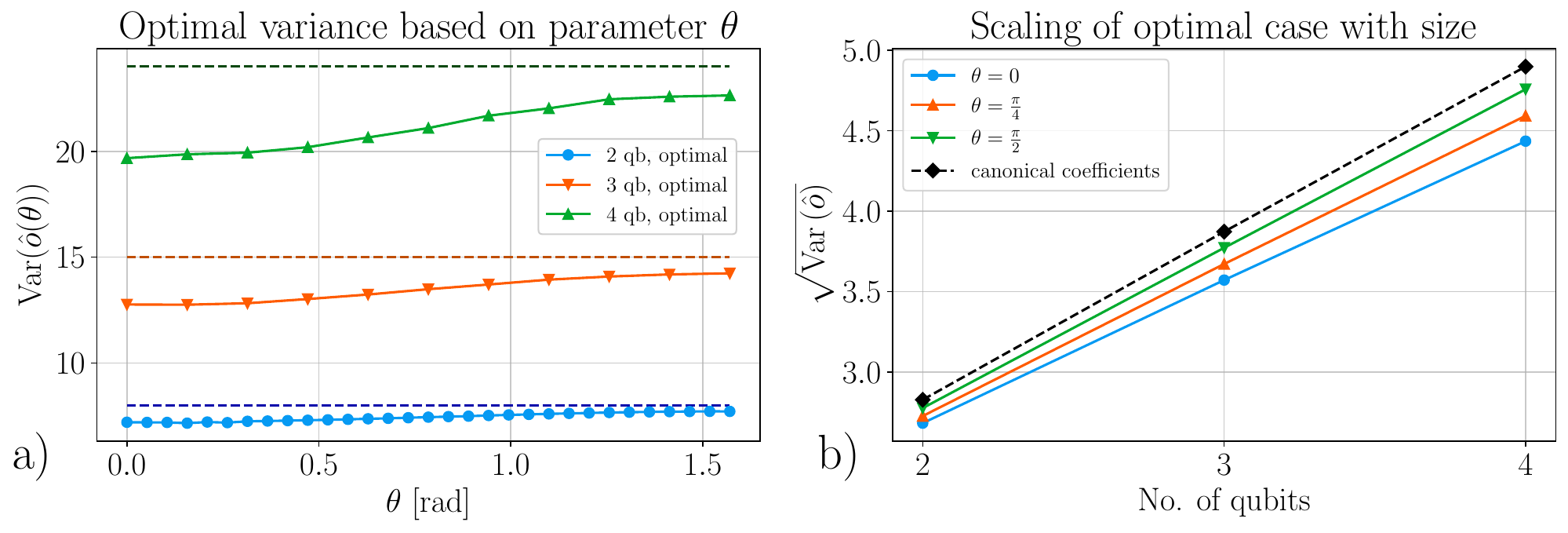}
\caption{ 
\textbf{Optimized variance of sums of parametrized local Pauli observables} 
The left panel a) shows the comparison of the optimal upper bound on variance (full line) with the canonical maximal variance (dotted line) in terms of the balance parameter $\theta$ for different number of qubits.
The right panel b) shows the scaling of the square root of the optimal variance (in order to highlight the scaling) with the number of qubits for chosen fixed values of $\theta$.
Black dotted line shows the scaling using canonical coefficients (the same  for every value of $\theta$ considered).
}
\label{fig:sum_pauli_results}
\end{figure*}

The optimisation method also yields results in more relevant situations, as for example, the estimation of the expectation value of an external magnetic field for interacting spin systems.
The observables considered therefore are defined as
\begin{equation}\label{eq:sum_pauli}
        O^{(N)}(\theta) 
        = \sum_{i=1}^N \hat{\sigma}_i(\theta), 
\end{equation}
where each term is defined as
\begin{equation}\label{eq:pauli_single_tens}
    \hat{\sigma}_i(\theta) = \Id^{\otimes (i -1)} \otimes \hat{\sigma}(\theta) \otimes \Id^{\otimes (N - i)},
\end{equation}
having the index $i$ indicate the position of the non-trivial parametrized combination of single-qubit Pauli operators
\begin{equation}\label{eq:theta_pauli}
    \hat{\sigma}(\theta) = \cos{\frac{\theta}{2}}  X + \sin{\frac{\theta}{2}} Z. 
\end{equation}
As before, due to the symmetry of the system, we consider the interval $\theta \in [0,\pi/2]$.

The optimization results in terms of the parameter can be found in Fig.\ref{fig:sum_pauli_results} a). 
The canonical estimator maintains a constant upper bound for all values of $\theta$, where the canonical inversion coefficients of $O^{(N)}(\theta)=\sum_{i_1\dots i_N}c^{(N)}_{i_1\dots i_N}E_{i_1\dots i_N}$ for $N$ qubits correspond directly to the linear composition 
\begin{equation}\label{eq:can_est_pauli_sum}
    \begin{split}
    \vec{c}^{\,(N)} & = \sum_{i=1}^N \vec{c}_i \\
                 & = \sum_{i=1}^N 
                \Id^{\otimes(i-1)}
                \otimes
                \vec{c}\otimes\Id^{\otimes(N-i)}
    \end{split}
\end{equation}
of the tensor extension of local canonical coefficients of the corresponding single-qubit observable $\sigma(\theta)=\sum_k c_k E_k$.
For Pauli observables, the canonical coefficients have been shown to also be optimal: the estimator built with the coefficients from \eqref{eq:can_est_pauli_sum} is equivalent then to first optimizing the local observables and combining the resulting optimal coefficients to match the same structure of the target observable, as introduced in \eqref{eq:sum_pauli}.

Nevertheless, executing the optimization protocol on the full global post-processing space results in optimal coefficients \( c^{(N)}_{i_1\dots i_N} \).
As in the previous example, we notice how the optimal coefficients cannot be decomposed in a linear structure as the canonical choice~\eqref{eq:can_est_pauli_sum}, rather are of a global or entangled nature (in the sense of decomposition as in ~\eqref{eq:local coeff}).
The scaling of the sampling complexity, as shown in Fig.\ref{fig:sum_pauli_results} b), we see the growth appears to be sub-exponential, with a slope dependence on the parameter $\theta$: initial analysis suggests a quadratic scaling with the number of qubits, as for the canonical scaling. 

\section{Discussion and outlook}\label{sec:outlook}
In the context of linear quantum tomography, once a non-adaptive POVM is fixed and measurement data are collected, there still lies freedom in classical post-processing: the same observable typically admits many unbiased linear reconstructions whenever the POVM is informationally overcomplete. 
In this work we made that freedom operational by casting the design of reconstruction coefficients as a convex minimax problem—minimizing the worst-case estimator variance over all states consistent with quantum mechanics, subject to unbiasedness. 
This yields an efficiently computable procedure with guaranteed convergence, and returns the tightest state-independent variance bound achievable by unbiased linear post-processing for the chosen POVM–observable pair. 
In this sense, it refines standard shadow-norm guarantees and turns “ask questions later” into “process later, but optimally,” without changing the experimental layer.

Our numerical examples show that the impact of variance-optimal post-processing can be substantial: beyond improving constants, it can lead to pronounced separations from canonical reconstructions and, for structured families of targets, can even alter the observed scaling with system size. 
This highlights that a significant portion of the sampling overhead commonly attributed to non-adaptive shadows may be a consequence of suboptimal reconstruction rather than a fundamental limitation of the measurement scheme itself.

Several questions follow naturally. 
First, what is the structure of the saddle point — i.e., the worst-case state that attains the optimized bound?
Classical extremal-variance principles suggest that “two-point” extremizers are typical; an attractive quantum analogue is that low-rank states (possibly rank two) might be sufficient to attain the worst-case variance for broad classes of POVMs and observables. 
Establishing such structural results would enable analytic bounds (or controlled approximations) and reduce reliance on full numerical maximization over density operators.

Second, when do global (non-factorizing) reconstruction coefficients provide a genuine scaling advantage, even when both the POVM and target observable are built from local tensor products? Pinning down conditions under which optimal coefficients must depart from product structure would clarify which improvements are “generic” and which rely on special symmetries or degeneracies. This also motivates developing constrained or structured variants of the optimization —e.g., enforcing locality, symmetry reduction, or tensor parameterizations of the coefficients— to obtain scalable algorithms with quantifiable performance loss relative to the fully optimal solution.

Finally, the framework invites extensions beyond single-observable optimization. 
One direction is optimization of nonlinear properties of the quantum state -e.g., purity, second-order Rényi entropy, two-point correlation functions. 
Another is robustness: incorporating known experimental imperfections (noise in the POVM, calibration uncertainty, finite sampling effects used to tune the estimator) directly into a robust minimax formulation. 
Together, these directions would strengthen the role of post-processing optimization as a practical, drop-in upgrade to shadow-based characterization pipelines.

\section*{Acknowledgments}
This research was funded in whole, or in part, by the Austrian Science Fund (FWF) (Grant No. 10.55776/F71) (BeyondC).

\bibliography{bibliography}

\appendix

\section{Scaling of sample size}\label{app:sample_size}
In this section we briefly discuss the statistical results which determine the sufficient sample size~\eqref{eq:sample_size} to achieve a $\varepsilon$-accuracy of the estimation.
As presented in the main text, the classical estimator of the expectation value of the target observable $A$ is built by associating to each POVM measurement outcome an estimator from the pre-determined set $\{\eta_j\}_j^{n}$ such that it forms a dual frame as defined in \eqref{eq:dual_frame}.
Due to this property, the estimator for the observable $O$
\begin{equation}\label{eq:classical_estimator}
    \hat{o}(K) = \frac{1}{K} \sum_{i=1}^K \Tr{A\,\eta^{(i)}}= \frac{1}{K} \sum_{i=1}^K x^{(i)}
\end{equation}
where $K$ represents the number of measurements, $x$ are the corresponding decomposition coefficient and the index $i$ indicates the particular measurement outcome.
Since $\{\eta_j\}\_j^{n}$ represent a dual frame, the estimator is guaranteed to have expectation value $\mathbb{E}[\hat{o}] = \Tr{O\,\rho}$.

The convergence to the true value can be controlled by classical concentration arguments, such as \textit{Hoeffdings's inequality}~\cite{inigo2021}: 
given the set $X_1,\ldots X_K$ of $K$ independent bounded variables, with $a_i \leq X_i \leq b_i$, the probability that the sum $S_K = \sum_i X_i $ deviates from its average can be bounded for any $t>0$ by
\begin{equation}\label{eq:hoeffding}
    \mathbb{P}(|\hat{S}_K-\mathbb{E}[\hat{S}_K]|>t) \leq 2\,\exp{\dfrac{-2 \,t^2}{\sum_{i=1}^K (b_i-a_i)^2}}.
\end{equation}
In the classical shadows case we can take as independent random variables 
\begin{equation}
    X_i = \dfrac{\hat{o}^{(i)}} {K} 
    = \dfrac{\Tr{O\,\eta^{(i)}}} {K},
\end{equation}
such that $S_K = \hat{o}(K)$ as introduced in \eqref{eq:classical_estimator}; as already discussed, $\mathbb{E}[\hat{o}] = \ev{O}$.
Given the upper bound on variance $\sigma^2 >\Var{\hat{o}}$, we can bound each independent variable as
\begin{equation}\label{eq:low_up_bounds}
    a_i = \dfrac{\ev{O} - \sigma}{K} 
    < X_i < 
    \dfrac{\ev{O} + \sigma}{K} = b_i : 
\end{equation}
$\forall 1\leq i \leq K$ we have $(b_i - a_i)^2 = 4\sigma^2/K^2$.
By substituting everything in \eqref{eq:hoeffding}, we see how the probability of the shadow estimator to deviate from the actual expectation value for more than an accuracy $\varepsilon$ decreases exponentially with the sample size $K$:
\begin{equation}\label{eq:hoeff_cs}
    \mathbb{P}(|\hat{o}-\ev{O}|>\varepsilon) \leq 2\,\exp{\dfrac{-K^2 \,\varepsilon^2}{2\sigma^2}}.
\end{equation}

By fixing the accuracy $\varepsilon$ with probability of success at least $1-\delta$, we may compute an estimate of the required sample size by inverting \eqref{eq:hoeff_cs}
\begin{equation}\label{eq:sample_size}
    K^2 > \dfrac{2\sigma^2}{\varepsilon^2} \log\frac{2}{\delta}. 
\end{equation}

Therefore minimization of the value of $\sigma$ guarantees an upper bound on $\Var{\hat{o}}$ for any sampled distribution so reduces the necessary sample size.

Given the flexibility with which classical shadows handles the measurement data to reconstruct linear properties of the system, the same outcomes can be used to estimate multiple properties simultaneously.

The increase of the required sample size can be estimated by relying on the \textit{union bound} ~\cite{inigo2021}:
given a countable series of events $V_1,V_2\ldots$ the probability of each taking place simultaneously is bounded by
\begin{equation}\label{eq:boole_ineq}
    \mathbb{P}\left(\cup_{i} V_i\right) \leq \sum_i \mathbb{P}(V_i).
\end{equation}
In our case, the events considered are $V_i = |\hat{o}_i-\ev{O_i}| > \varepsilon$. 
The requirement of a maximum tolerance $\delta$ on the total probability of each of the estimators deviating for more that $\varepsilon$ can then be simply achieved by bounding each single probability with $\frac{\delta}{M}$.
From \eqref{eq:hoeff_cs} this implies that the sample size in \eqref{eq:sample_size} gains a scaling factor proportional to $\log M$ ~\cite{preskill2020,dakic2022}:
\begin{equation}\label{eq:sample_size_final}
    K^2 > \dfrac{2\sigma^2}{\varepsilon^2} \log\frac{2M}{\delta}. 
\end{equation}

It can be noted how \eqref{eq:sample_size_final} actually slightly overestimates the required sample size, since \eqref{eq:low_up_bounds} and \eqref{eq:boole_ineq} are not actually saturated in the cases considered.
However the actual improvement which can be achieved is so slight that it is massively overshadowed by any other contribution to the error and is not really worth analyzing.
The relevance of this kind of sample size analysis is found in determining the dimensional scaling achievable rather than the tightest estimation possible of the sample size.

\section{Convexity and concavity of variance for fixed terms}\label{app:conc_conv}
In this section we characterize the convexity of the bilinear function
\begin{equation}\label{eq:var_bilinear}
    f(\vec{x},\rho) = \sum_{j=1}^n |x_j|^2 \Tr{\rho\,E_j} - \left(\sum_{j=1}^n x_j^2 \Tr{\rho\,E_j}\right)^2,
\end{equation}
with the constraints on $\vec{x}$ to reconstruct the target observable 
\begin{equation}\label{eq:x_constraint}
    \sum_{j=1}^n x_j E_j = O,
\end{equation}
and on $\rho$ to be a valid density matrix,
\begin{equation}\label{eq:rho_constraints}
    \Tr{\rho} = 1, \quad \rho \succ 0   
\end{equation}
In particular, we show that $f(\vec{x},\cdot) : \mathbb{B}\to\mathbb{R}$ for fixed state is convex in the variable $\vec{x}$ while, for fixed set of coefficients, the function $f(\cdot,\rho) : \mathcal{R}\to\mathbb{R}$ is concave.

\subsection{Convexity for fixed state}\label{app:conv}
Convexity of $f$ requires that, given two valid solutions $\vec{x}$ and $\vec{y}$ which satisfy the constraints from \eqref{eq:x_constraint}, then
\begin{equation}\label{eq:convexity_req}
    f(\lambda \,x + (1-\lambda)\,y,\bar{\rho}) \leq \lambda\,f(x,\bar{\rho}) + (1-\lambda)\,f(y,\bar{\rho}) 
\end{equation}
for all fixed states $\bar{\rho}$ and all values of $\lambda \in [0,1]$.
The left-hand side can be expanded into
\begin{equation}\label{eq:conv_lhs}
    \begin{split}
        f(\lambda \,x& + (1-\lambda)\,y,\,\bar{\rho}) \\
        = &  
        \sum_{j=1}^n |\lambda \,x_j + (1-\lambda)\,y_j|^2\,\Tr{\bar{\rho}\,E_j} \\
        & - 
        \left(\sum_j (\lambda \,x_j + (1-\lambda)\,y_j)\Tr{\rho\,E_j}\right)^2, 
    \end{split}
\end{equation}
while the right-hand side turns into
\begin{equation}\label{eq:conv_rhs}
    \begin{split}
        \lambda\,&f(x,\bar{\rho}) + (1-\lambda)\,f(y,\bar{\rho})  \\
        =  &
                \lambda \sum_j |x_j|^2\Tr{\bar{\rho}\,E_j} +
                (1-\lambda) \sum_j|y_j|^2\Tr{\bar{\rho}\,E_j}
           \\
         & - 
                \lambda\,\sum_j x_j\Tr{\bar{\rho}\,E_j})^2 
                +(1-\lambda)\,(\sum_j y_j\Tr{\bar{\rho}\,E_j})^2 
    \end{split}
\end{equation}

By using the constraint \eqref{eq:x_constraint}, the second term in both \eqref{eq:conv_lhs} and \eqref{eq:conv_rhs} cancel out, as they both amount to the expectation value of the target observable:
\begin{equation}\label{eq:same_ev}
\begin{split}
    \left(\Tr{\bar{\rho}\,O}\right)^2
    &= \lambda\,\left(\Tr{\bar{\rho}\,O}\right)^2 
    +(1-\lambda)\,\left(\Tr{\bar{\rho}\,O}\right)^2 \\
    &= \left(\lambda \Tr{\bar{\rho}\,O} + (1-\lambda) \Tr{\bar{\rho}\,O}\right)^2
    \end{split}
\end{equation}
for any value of $\lambda$.

Therefore the convexity can be determine by only looking a the first (quadratic) term of the variance.
The inequality \eqref{eq:convexity_req} reduces to
\begin{equation}\label{eq:ineq_conv}
    \begin{split}
     \lambda^2 &\sum_j |x_j|^2\Tr{\bar{\rho}\,E_j}
        (1-\lambda)^2 \sum_j |y_j|^2\Tr{\bar{\rho}\,E_j} \\
         &+ 2\,\lambda\,(1-\lambda) \sum_j |x_j||y_j|\Tr{\bar{\rho}\,E_j} \\
        \leq &
        \lambda \sum_j |x_j|^2\Tr{\bar{\rho}\,E_j} +
        (1-\lambda) \sum_j|y_j|^2\Tr{\bar{\rho}\,E_j},
    \end{split}
\end{equation}
which can be reordered into
\begin{equation}\label{eq:ineq_conv}
    \begin{split}
     \lambda&(1-\lambda)\,
        \sum_j |x_j|^2\Tr{\bar{\rho}\,E_j}\\
        &+\lambda(1-\lambda)\,\sum_j |y_j|^2\Tr{\bar{\rho}\,E_j}\\
        &- 2\,\lambda(1-\lambda)\, \sum_j |x_j||y_j|\Tr{\bar{\rho}\,E_j} \\
    = &
     \lambda(1-\lambda) \sum_{j=1}^n |x_j-y_j|^2 \;\Tr{\bar{\rho}\,E_j} \geq 0.
    \end{split}
\end{equation}
Since $\bar{\rho}$ satisfies \eqref{eq:rho_constraints}, each term $p_j = \Tr{\bar{\rho}\,E_j} \geq 0$, the inequality is always satisfied $\forall \lambda \in[0,1]$; equality is found only for the extremes $\lambda \in \{0,1\}$, thus variance for fixed states is strictly convex.

\subsection{Concavity for fixed set of coefficients}
The concavity of the bilinear variance function from  for fixed set of coefficients $\bar{x}$ can be similarly shown by treating separately the two different terms in \eqref{eq:var_bilinear}.

The first term 
\begin{equation}\label{eq:first_term_var}
    \sum_{j=1}^n |\bar{x}_j|^2 \;\Tr{\rho\,E_j}
\end{equation}
is obviously linear in the density matrix, therefore it is both concave and convex.
Similarly, it is simple to see that the second term
\begin{equation}\label{eq:second_term_var}
    \left(\sum_{j=1}^n \bar{x}_j^2 \;\Tr{\rho\,E_j}\right)^2
\end{equation}
is instead quadratic and therefore, as shown explicitly in the previous subsection, strictly convex.
However, since in \eqref{eq:var_bilinear} this term is subtracted, for any fixed set of coefficients the function $f(\bar{x}, \rho)$ is strictly concave overall.

\section{Minimal variance coefficients}\label{app:opt_coef}
In this section we explicitly show the calculations that lead to the expression of the optimal set of coefficients for fixed states, as introduced in \eqref{eq:opt_coef_state}.

We first consider the minimization problem for a fixed density matrix as introduced in \eqref{eq:min_opt}.
We introduce the $n\times n$ diagonal probability distribution matrix
\begin{equation}\label{eq:diag_prob_map}
    P(\bar{\rho}) = \bar{P} = \diag{(p_j)}= \diag{(\Tr{\bar{\rho}\,E_j})},
\end{equation}
defined by the fixed density matrix $\bar{\rho}$.
Since the matrix $\bar{P}$ is diagonal by construction, the variance can also be expressed as
\begin{equation}\label{eq:var_alt_expression}
    \Var{\hat{o}\,\big|\,\vec{x},\bar{\rho}} = \vec{x}^\dagger \bar{P} \vec{x} - \vec{x}^\dagger \bar{P}^{\,\dagger} \bar{P} \vec{x}.
\end{equation}
As also discussed in Section \ref{sec:var_min}, and in Appendix \ref{app:conv}, the target function can be simplified to a quadratic form.

As already mentioned in Section \ref{sec:var_min}, the $n\times D$ coefficient matrix $R$ for the POVM is defined such that each entry is the projection of a POVM effect onto the element of an orthonormal basis $\{B_k\}_{k=1}^D$ of $\mathcal{V}_D$: 
\begin{equation}\label{eq:r_def}
	R_{jk} = \ev{E_j,B_k} = \Tr{E_j B_k^\dagger}.
\end{equation}
Each row contains the coefficients to reconstruct the POVM effects in terms of the basis elements: $E_j = \sum_k R_{jk} B_j$.
By also decomposing the target observable $O$ in terms of the same basis, the constraint on coefficients $\vec{x}$ can be expressed as an affine expression:
\begin{align}
	O &= \sum_{j=1}^n x_j E_j = \sum_{j=1}^n x_j  \sum_{k=1}^D R_{jk} B_j \label{eq:O_coef_constr}\\
	 & = \sum_{k=1}^D \ev{O,B_k} B_k \label{eq:O_coef_basis} \\
	 \Rightarrow &\quad R^\top\vec{x}  = \vec{o} \label{eq:O_aff_constr} 
\end{align}
By also introducing the POVM coefficient matrix as in \eqref{eq:r_def}, we can express the minimization problem~\eqref{eq:min_opt} as a Quadratic Problem, which is guaranteed to yield a solution, when available, in polynomial time:
\begin{equation}\label{eq:QP_program}
    \begin{split}
        \min_{x_j \in \mathbb{B}}  \quad 
        & \vec{x}^\dagger \, \bar{P}\,\vec {x} \\
        \text{s.t.} \quad
        & R^\top \vec{x} = \vec{o}.
    \end{split}
\end{equation}

In the case in which the matrix $\bar{P}$ is invertible, that is $\forall j \Tr{\bar{\rho}\,E_j} = \bar{p}_j \neq 0$,
an analytical solution can be determined by finding the extremal of the Lagrangian
\begin{equation}
	\mathcal{L}(\vec{x},\vec{\lambda}) = \vec{x}^\dagger \,\bar{P}\,\vec x - \vec{\lambda}^\top (\, R^\top\vec{x} - \vec{o}\,),
\end{equation}
where $\vec{\lambda}$ represent the Langrange multipliers associated to each equality constraint.
The extremal can be found by fixing the gradient of $\mathcal{L}$ to zero:
\begin{numcases}{\grad\mathcal{L}(\vec{x},\vec{\lambda})= }
    \pdv{\mathcal{L}}{x_j} 
    &$= 2 \bar{p}_j x_j - \vec{\lambda}^\top R^\top
    = 0$ \label{eq:lag_der_x}\\
    \pdv{\mathcal{L}}{\lambda_k} 
    &$= \sum_{j=1}^n R_{kj}^\top x_j - o_k 
    = 0.$ \label{eq:lag_der_l}
\end{numcases}

If $\bar{p}_j \neq 0$, from \eqref{eq:lag_der_x} we find an easy expression for $x_j$ in terms of the Lagrange multipliers as
\begin{equation}\label{eq:x_sol_lambda}
    x_j = \frac{1}{2} \,\frac{1}{\bar{p}_j}\, \sum_k R_{jk} \,\lambda_j. 
\end{equation}
This expression can be inserted in \eqref{eq:lag_der_l} to find a system of equations in which, together with the Lagrange multipliers, only the POVM coefficient matrix and the target observable coefficients appear.
This system of equations can be condensed in a vector expression
\begin{equation}\label{eq:lambda_sol}
   \frac{1}{2}\, R^\top \bar{P}^{-1} R \,\vec{\lambda} = Q \, \vec{\lambda} = \vec{o},
\end{equation}
where we introduce the $D \times D$ square matrix $Q$.
Since, by definition, $R$ is full rank $r=D$, as is $\bar{P}$\footnote{with the assumption $p_j\neq 0$}, the square matrix $Q$ is also full rank and therefore always invertible.
This returns a closed expression for $\vec{\lambda}$
\begin{equation}\label{eq:lambda_opt}
    \vec{\lambda} = 2\, Q^{-1} \,\vec{o},
\end{equation}
which, when inserted in \eqref{eq:x_sol_lambda}, returns exactly the optimal result already anticipated in \eqref{eq:opt_coef_state}.

The assumption $p_j\neq0 \forall j$, required for the analytical solution, is, in general, non-trivial: while we are guaranteed that $\Tr{\rho\,E_j}>0$ for at least one $j$ if $\rho\in\mathcal{V}_D$ , there might very well be situations in which for some $k$ $\Tr{\rho\,E_k}=0$.
Notable examples include also the cases considered in this work, as for example $\dyad{0}{0}$ is always orthogonal to the effects of the plane POVM from \eqref{eq:flat_povm_def} for $\zeta=x$.
In the case in which the diagonal matrix $P(\rho)$ is not full rank, then it cannot be inverted and the solution here presented does not hold.
Moreover, while the QP in \eqref{eq:QP_program} is guaranteed to halt in polynomial time, it is not guaranteed to return a solution.
A simple method to ensure the invertibility of $\bar{P}$ is to act, for any POVM describing the system, with the infinitesimal perturbation
\begin{equation}\label{eq:eff_shift}
    E_j \mapsto \left(1-\varepsilon\right)\, E_j + \frac{\varepsilon}{n} \Id_D.
\end{equation}
For very small values $\varepsilon \ll 1$ we can keep the probability distribution within $\varepsilon-$accuracy, $p_j^{(\varepsilon)}\approx p_j$, yet with the guarantee $p_j^{(\varepsilon)} > \varepsilon/D>0\;\forall j$.
The shift itself is not simply a mathematical artifact but is physically motivated, as it corresponds to the inevitable presence of dephasing noise in the measurement scheme.
Therefore, whilst not universal, in all extent the solution from \eqref{eq:opt_coef_state} can be considered valid for most realistic cases, with the caveat of the consideration of a small dephasing noise in the description of the set-up.

As introduced in Section \ref{sec:var_min}, the optimal choice of coefficients $\vec{x}(O,\bar{\rho})$ can be expressed as the product of a state-dependent coefficient matrix $L^\star (\bar{\rho})$ and of the coefficients of the target observable, decomposed in the same basis. 
This matrix represents a valid pseudo-inverse of the rectangular coefficient matrix $R$, that is a matrix such that $R^\top\,L = \Id_D$. 
Using this $L^\star$ as a coefficient matrix to construct estimators $\eta^{\,\star}_j = \sum_k L^\star_{jk} B_j$, we reproduce the well-known result for the optimal dual frame for single-shot variance ~\cite{innocenti2023}.
The equivalence between optimizing the decomposition coefficients $\vec{x}$ and finding the optimal set of estimators $\{\eta_k\}_{k=1}^n$ therefore lies in finding the optimal pseudo-inverse $L^\star$ of the POVM coefficient matrix: the choice of pseudo-inverse determines a different set of valid dual frames, where the optimal set presents a direct dependence on the density matrix of the unknown state.

\section{Canonical pseudo-inverse and coefficients}\label{app:canonical_inv}
As mentioned in the previous section, any pseudo-inverse of the POVM coefficient matrix $R$ defines a valid set of estimators.
It is worth examining the case of the Moore-Penrose pseudo-inverse:
\begin{equation}\label{eq:mp_pseudoinverse}
    L^\star\left(\Id_D / D \right) 
    = L_{MP} 
    = R\,\left(R^\top\, R\right)^{-1}.
\end{equation}
This choice of pseudo-inverse determines the solution $\vec{x}_{MP}$ to the linear problem from~\eqref{eq:O_coef_constr} with the smallest Euclidean vector norm ~\cite{planitz1979}:
\begin{equation}\label{eq:std_coef}
    \vec{x}_{MP} = L_{MP} \,\vec{o} = \min_{R^\top \vec{x} = \vec{o}} \|\vec{x}\|_2.
\end{equation}
It can also be interpreted as the particular case of the state-dependent optimal solution ~\eqref{eq:opt_coef_state} defined by the maximally-mixed state.
The estimators obtained, and thus also decomposition coefficients $\vec{x}_{MP}$, correspond to the optimal choice when no additional assumptions are made, as $\rho = \frac{\Id_D}{D}$ represents the average unbiased state.
This implies that, without taking into consideration the actual measured distribution, this choice of coefficients minimizes the apriori average variance ~\cite{innocenti2023}.
For this reason, this solution is referred to as the ``canonical'' or "standard" choice of estimators, which coincides with the one first introduced with Classical Shadows ~\cite{preskill2020}.

The canonical choice of estimators represents the optimal choice for the estimation of the classical shadow without any assumption on the property to be estimated.
However, when estimating the expectation value of a particular observable, this choice does not guarantee a tight upper bound for any state in the subspace considered.

\section{Maximal variance density matrices}
Unlike the optimization problem introduced in \eqref{eq:min_opt}, the optimal solution for maximum variance and maximum shadow norm will be, in general, different.
This can be seen from the fact that the shadow norm $\|O\|^2_{\text{SN}} = \sum_j |x_j|^2 p_j$ is linear in $\vec{p}$, therefore it finds its maximum on the boundary of $\mathcal{R}$, that is the set of pure states.
More specifically, since the maximum of $\mathbb{E}[\hat{o}] = \max$ eigv $(O)^2$ for the optimal measurement case, we expect the maximal state to be the eigenstate of $O$ with corresponding largest eigenvalue (in absolute value); in case the spectrum of $O$ is degenerate, the search can be limited to the subspace $\mathcal{E}_{\max} \subset \mathcal{V}_D$ spanned by the eigenstates of $O$ with the same maximal eigenvalue.
Instead, the variance presents a concave dependence on $\vec{p}(\rho)$, therefore the optimal value is generally expected to be in the interior of $\mathcal{R}$. 

The problem of maximum variance of a Hermitian observable has a well-known solution~\cite{textor1978} which depends on the maximum and minimum eigenvalue of the observable:
\begin{equation}\label{eq:max_var_hermitian}
    \max \Var{O} = \frac{1}{4}\left(\lambda_{\max}-\lambda_{\min}\right)^2.
\end{equation}
However, as far as the authors are aware, there is little characterization of which quantum states actually achieve this variance.
On its own, \eqref{eq:max_var_hermitian} does represent the lowest possible sample size to guarantee $\varepsilon-$accuracy to the convergence of the estimator to the actual expectation value for any unknown underlying state, but it gives no information on how to actually build the estimator.
Therefore, numerical methods are required to determine the maximal state and thus the optimal choice of estimators.

Moreover, the direct maximization of the variance of the observable in terms of quantum states is misleading: the interest is to minimize the upper bound of the variance of the \textit{classical estimator} $\hat{o}$, which rather than depending directly on the observable it depends on its decomposition in terms of the POVM effects.
Thus the optimization function must also take into account not only the state which achieves maximal variance but also the corresponding optimal coefficients.
By using the analytical dependence of the coefficients from the density matrix~\eqref{eq:opt_coef_state} the target optimization function can be expressed in terms of a single variable~\eqref{eq:max_problem_states}.

\end{document}